# Rotation-translation coupling of a double-headed Brownian motor in a traveling-wave potential

Wei-Xia Wu[1], Chen-Pu Li[2], Yan-Li Song[3], Ying-Rong Han[4], and Zhi-Gang Zheng[5,†]

[1] *Science Education Department, Beijing Institute of Graphic Communication, Beijing 102600, China*
[2] *College of Science, Hebei University of Architecture, Zhangjiakou 075000, China*
[3] *School of science, Tianjin University, Tianjin 300072, China*
[4] *School of science, Hebei University of Technology, Tianjin 300401, China*
[5] *Institute of Systems Science and College of Information Science and Engineering, Huaqiao University, Xiamen 361021, China*
Corresponding author. E-mail: † zgzheng@hqu.edu.cn

Considering a double-headed Brownian motor moving with both translational and rotational degrees of freedom, we investigate the directed transport properties of the system in a traveling-wave potential. It is found that the traveling wave provides the essential condition of the directed transport for the system, and at an appropriate angular frequency, the positive current can be optimized. A general current reversal appears by modulating the angular frequency of the traveling wave, noise intensity, external driving force and the rod length. By transforming the dynamical equation in traveling-wave potential into that in a tilted potential, the mechanism of current reversal is analyzed. For both cases of Gaussian and Lévy noises, the currents show similar dependence on the parameters. Moreover, the current in the tilted potential shows a typical stochastic resonance effect. The external driving force has also a resonance-like effect on the current in the tilted potential. But the current in the traveling-wave potential exhibits the reverse behaviors of that in the tilted potential. Besides, the currents obviously depend on the stability index of the Lévy noise under certain conditions.

## I. INTRODUCTION

Brownian motors, also known as Brownian ratchets, can convert the nonequilibrium fluctuation into useful work to achieve the directed motion of themselves [1, 2]. Understanding the motion mechanism of Brownian motors is a fundamental issue to understand and describe the dynamical behavior of many biological or physical systems such as protein motors in cell and Josephson junctions. The motion mechanism of Brownian motors has attracted much interest in physics, biology and other fields for a few decades [3-15].

A large number of experimental and theoretical studies on Brownian motors have been carried out and a series of meaningful results have been obtained [16-23]. Inspired by the Smoluchowski–Feynman ratchet, various theoretical models for Brownian motors have been proposed. Among them, most of early models focused on a single Brownian motor which was regarded as a Brownian particle moving unidirectionally in a periodic potential [24, 25]. In recent years, however, experiments have revealed that multiple Brownian motors can work together to achieve specific functions. For instance, protein motors such as kinesins, myosins and dyneins, often cooperate to carry out the intracellular transport. Moreover, a majority of protein motors have two identical motor domains, called "heads", which coordinate each other to achieve a 'hand-over-hand' walk along linear tracks [26-30]. So far, great attention has been paid to coordinated transport, especially the directed transport of double-headed Brownian motors. Due to the directed motion of a double-headed Brownian motor along a periodic potential, only the translational degree of freedom has been concerned in most of works [31-34]. Recently, the development of experimental technology has brought more exciting findings. Howard suggested that a kinesin rotates about an axis perpendicular to the microtubule by biochemical experiment [35]. Toprak et al. observed that there is a rotational motion when myosin V walks along the actin by fluorescence imaging with one-nanometer accuracy [36]. Dunn and his co-worker also found experimentally that the unbounded head of myosin V rotates freely about the lever-arm junction as it is released from the actin filament [37]. These findings showed that there is not only a translational degree of freedom but also a rotational degree of freedom when a double-headed Brownian motor moves along a periodic potential. Up to date, a few studies have given attention to this issue. B. Geislinger and his co-worker investigated three models of Brownian motors driven by rotation-translation coupling in a flashing potential, and analyzed the rotation-translation coupling mechanism [38]. Qiao et al. explored the rotational effect of a coupled dimer in a two-dimensional asymmetric periodic potential [39].

Moreover, it is well known that various random fluctuations exist inevitably in the surrounding environment when a system works. These random fluctuations can be modeled by noises. Several interesting noise-induced effects have been found in many systems like stochastic resonance, resonant activation and noise enhanced stability [40-46]. Similarly, when a Brownian motor moves in a periodic potential, there are various random fluctuations such as the thermal fluctuation. Gaussian noises are commonly selected to model the thermal fluctuation [47-50]. In a Gaussian distribution the tails of the probability density fall off exponentially, and the mean squared displacement in the passive motion grows linearly with time. Besides, non-Gaussian noises have also been widely concerned such as Lévy noises [51, 52]. In contrast to Gaussian noises, Lévy noises have long tailed distributions that decay as power laws. Some effects induced by Gaussian or non-Gaussian noises have also been revealed in Brownian motor systems [53-57].

In this work, we propose a theoretical model of a double-headed Brownian motor in which both the translational and rotational degrees of freedom are concerned. The directed transport properties of the motor system in a traveling-wave potential are investigated. The effects of various parameters on the directed current in the translational direction are discussed, including the angular frequency of the traveling wave, noise intensity, external driving force and the rod length. We firstly focus on Gaussian white noise in the rotational direction, and then white stable Lévy noise. The transport behaviors under these two types of noises are compared.

## II. ROTATION-TRANSLATION COUPLING MODEL

We consider a double-headed Brownian motor system consisting of two identical heads which are connected by a rigid rod with the length $l$ in a traveling-wave potential, which is shown in Fig.1. The traveling-wave potential is chosen as the harmonic wave with the amplitude $U_0$, wavelength $\lambda$ and angular frequency $\omega$

$$U_x(x,t) = U_0 \cos(\frac{2\pi}{\lambda}x - \omega t). \tag{1}$$

The wave velocity u is related to $\omega$ and $\lambda$ through the formula $u = \omega\lambda/2\pi$. If the value of $\lambda$ is fixed, $u$ is only determined by $\omega$. We further consider the rotational degree of freedom of the double-headed motor, and suppose the motor to be in a periodic potential in the rotational direction as follows

$$U_\theta(\theta) = d\cos\theta, \tag{2}$$

where $\theta(t)$ is the rotation angle of the motor around the center at time $t$, and $d$ is the amplitude of the periodic potential. Then the total potential can be written as

$$U_t = U_x(x,t) + U_\theta(\theta). \tag{3}$$

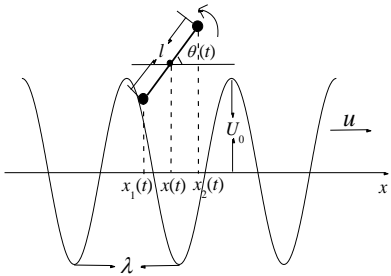

Fig. 1. Schematic diagram of a double-headed Brownian motor in a traveling-wave potential $U_x(x,t)$.

The overdamped dynamics of the motor system can be given by the following equations

$$\gamma_1 \dot{x}_i = -\frac{\partial U_t}{\partial x_i}, \tag{4}$$

$$\gamma_2 \dot{\theta} = F - \frac{\partial U_t}{\partial \theta} + \xi(t), \tag{5}$$

where $x_i$ ($i=1, 2$) is the translational position of the $i$-th head in the traveling-wave potential. $\gamma_1$ and $\gamma_2$ are the friction coefficients in the translational and rotational directions, respectively. $F$ is an external constant driving force in the rotational direction which can induce a continuous rotation of the double-headed Brownian motor. $\xi(t)$ denotes the noise from the nonequilibrium fluctuation which is subject to the system in the rotational direction. It is first considered as Gaussian white noise with $\langle\xi(t)\rangle = 0$ and $\langle\xi(t)\xi(t')\rangle = 2D\delta(t-t')$, where $D$ is the noise intensity.

Assuming that $x$ is the center-of-mass position in the translational direction, the positions of two motor heads can be written as

$$x_1 = x - \frac{l}{2}\cos\theta, \; x_2 = x + \frac{l}{2}\cos\theta. \tag{6}$$

Then the center-of-mass velocity can be expressed by

$$\dot{x} = \frac{1}{2}(\dot{x}_1 + \dot{x}_2). \tag{7}$$

By solving Eqs. (1), (3), (4), (6) and (7), we obtain specifically the overdamped dynamics of the center of mass as

$$\gamma_1\dot{x} = \frac{2\pi U_0}{\lambda}\cos(\frac{\pi a}{\lambda}\cos\theta)\sin(\frac{2\pi}{\lambda}x - \omega t). \tag{8}$$

For simplicity, we only use dimensionless units. So, by combining Eqs. (2), (3), (5) and (8), we obtain the following rotation-translation coupling model of the double-headed Brownian motor as

$$\dot{x} = \frac{2\pi U_0}{\lambda}\cos(\frac{\pi a}{\lambda}\cos\theta)\sin(\frac{2\pi}{\lambda}x - \omega t), \tag{9}$$

$$\dot{\theta} = F - d\sin\theta + \xi(t). \tag{10}$$

Equations (9) and (10) describe the translational motion of the center of mass and the rotational motion of the motor system, respectively. Here, the double-headed motor system given in Eq. (10) is regarded as a simple pendulum that can freely rotate counterclockwise and swing at a large angle.

We numerically simulate Eqs. (9) and (10) by using the stochastic Runge–Kutta algorithm. Each trajectory evolves $1\times10^5$ steps with the time step $\Delta t = 10^{-3}$. In the following calculations, some necessary parameters are set to be $U_0=1.0$, $\lambda=1.0$ and $d=1.0$. In this work, we are mainly concerned with the translational motion of the double-headed Brownian motor. So the directed current in the traveling-wave potential is defined by

$$\langle\dot{x}\rangle = \lim_{T\to\infty}\frac{1}{T}\int_0^T \dot{x}(t)dt \tag{11}$$

## III. RESULTS AND DISCUSSION

### A. TIME EVOLUTION OF THE POSITIONS

First, we investigate the time evolution of two head positions in the translational direction for different angular frequencies of the traveling wave in Fig. 2. It is shown that for a small angular frequency $\omega=1.0$ (i.e., small wave velocity), the motor moves in the negative direction, namely the opposite direction of the traveling wave in Fig. 2(a). While for a large angular frequency $\omega=10.0$ (i.e., large wave velocity), the motor moves in the positive direction in Fig. 2(b). It indicates that for different angular frequencies or wave velocities, the movement direction of the motor is different. With increasing the angular frequency (or the wave velocity), the current is possibly reversed. Moreover, in Figs. 2(a) and (b), we can clearly find that the positions of two motor heads change alternately with time, namely during one step one head leads and the other head leads during next step. This indicates that the motor walks in a hand-over-hand manner along the potential.

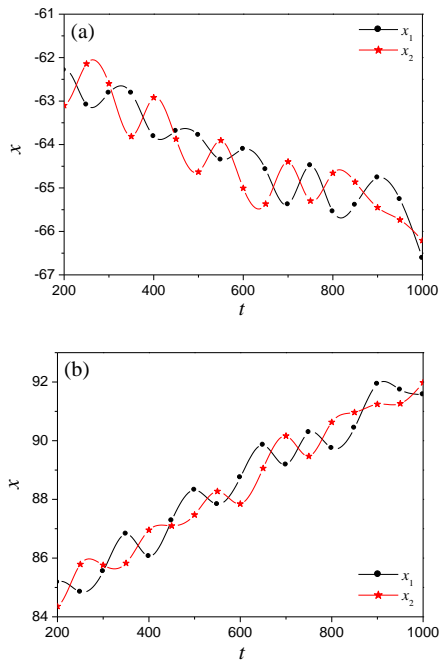

Fig. 2 The positions of the two motor heads in the traveling-wave potential versus time for (a) a small angular frequency $\omega=1.0$ (i.e., small wave velocity) and (b) a large angular frequency $\omega=10.0$ (i.e., large wave velocity). Parameter values: $l=1.0$, $D=1.0$, $F=5.0$.

In order to explain the result of the different movement directions for different angular frequencies of the traveling wave, next we further analyze the motion of the double-headed Brownian motor by transformation equation [1, 58]. First, Eq. (1) can be rewritten as

$$U_x(x,t) = U_0 \cos\left[\frac{2\pi}{\lambda}(x-ut)\right], \quad (12)$$

where $u$ is the wave velocity. We introduce a new space variable $y$ and suppose $y = x - ut$. Then

$$\langle \dot{x} \rangle = \langle \dot{y} \rangle + u. \quad (13)$$

From Eq. (13), one can find that the current in $x$ variable space should be equal to the sum of the current in $y$ variable space and the wave velocity. Eq. (4) can be reformulated as

$$\gamma_1 \dot{y}_i = -\frac{\partial U'_t}{\partial y_i}, \quad (14)$$

where $y_i$ ($i=1, 2$) is the translational position of the $i$-th head in $y$ variable space. $U'_t$ is the total potential in $y$ variable space

$$U'_t = U_y(y) + U_\theta(\theta), \quad (15)$$

where $U_y(y)$ is the effective potential in translational direction in $y$ variable space which is written as

$$U_y(y) = U_0 \cos\left(\frac{2\pi}{\lambda} y\right) + uy. \quad (16)$$

The potential in the rotational direction $U_\theta(\theta)$ is still the same as that in Eq. (2). In Eq. (16), we can see that the traveling-wave potential is equivalent to a tilted potential as shown in Fig. 3.

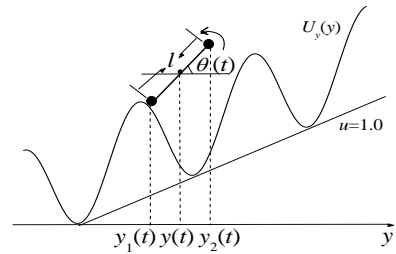

Fig. 3. Schematic diagram of the double-headed Brownian motor in an equivalent tilted potential $U_y(y)$.

The potential in the rotational direction $U_\theta(\theta)$ is still the same as that in Eq. (2). In Eq. (16), we can see that the traveling-wave potential is equivalent to a tilted potential as shown in Fig. 3. The wave velocity u denotes the tilt of the potential. We deal with the variable y in the same way as $x$ and obtain the rotation-translation coupling model of the double-headed Brownian motor in the tilted potential by combing Eq. (10) as follows

$$\dot{y} = \frac{2\pi U_0}{\lambda} \cos\left(\frac{\pi a}{\lambda} \cos\theta\right) \sin\left(\frac{2\pi}{\lambda} y\right) - u, \quad (17)$$

$$\dot{\theta} = F - d\sin\theta + \xi(t). \quad (18)$$

The directed current in the tilted potential is defined by

$$\langle \dot{y} \rangle = \lim_{T \to \infty} \frac{1}{T} \int_0^T \dot{y}(t)dt. \quad (19)$$

According to Eq. (13), the current $\langle \dot{x} \rangle$ in the traveling-wave potential should be the sum of the current $\langle \dot{y} \rangle$ in the tilted potential and the wave velocity $u$. In fact, it can be understood that two currents $\langle \dot{x} \rangle$ and $\langle \dot{y} \rangle$ are corresponding to two different reference frames respectively. $\langle \dot{x} \rangle$ denotes the current observed in one frame S fixed on the ground, called fixed frame. $\langle \dot{y} \rangle$ represents the current observed in the other frame S' fixed on the traveling wave, called mobile frame. The wave velocity $u$ is the velocity of the mobile frame S' relative to the fixed frame $S$. Therefore, Equation (13) indicates the well-known Galilean transformation. On the other hand, if observed in the fixed frame $S$, the center of mass participates in two translational motions. One is induced by the rotational motion, and the velocity is $\langle \dot{y} \rangle$. The other is the motion following the traveling wave, and the velocity is the wave velocity $u$. $\langle \dot{x} \rangle$ is just the total velocity. Furthermore, we can predict that the current $\langle \dot{y} \rangle$ should be negative because it is much easier to move for the center of mass downward than upward along the slope of the potential as shown in Fig. 3. In other word, the direction of the center-of-mass motion induced by the rotational motion should be opposite to that of the traveling wave in the fixed frame $S$.

The time evolution of the two head positions in the tilted potential under the same conditions as that in Fig. 2 is plotted in Fig. 4. Comparing to Fig. 2, we can see that for both cases $\omega=1.0$ and $\omega=10.0$, the center of mass of the motor moves indeed in the negative direction of $y$ and has different velocities $\langle \dot{y} \rangle$ for different tilts of the potential. According to Eq. (13), when $\langle \dot{y} \rangle$ is negative and the absolute value $|\langle \dot{y} \rangle|$ is larger than the wave velocity $u$ ($u$ is always positive), the current $\langle \dot{x} \rangle$ is negative. On the contrary, $\langle \dot{x} \rangle$ is positive. Therefore, the direction of the directed current $\langle \dot{x} \rangle$ in the traveling-wave potential is possibly different for different angular frequencies $\omega$ (i.e., different wave velocities). It is implied that the current reversal in the traveling-wave potential is caused by the competition between the negative velocity of the center of mass induced by the rotational motion and the positive wave velocity, which is different from the mechanism of current reversal in other models [54, 59].

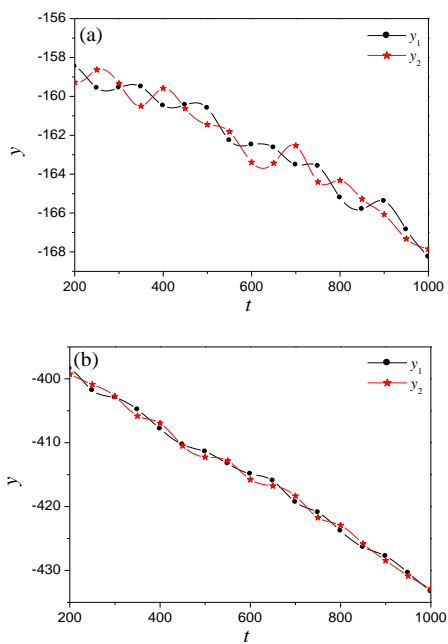

Fig. 4 The positions of the two motor heads in the tilted potential versus time for (a) a small angular frequency $\omega=1.0$ (i.e., small wave velocity) and (b) a large angular frequency $\omega=10.0$ (i.e., large wave velocity). Parameter values: $l=1.0$, $D=1.0$, $F=5.0$.

## B. EFFECT OF THE ANGULAR FREQUENCY ON THE DIRECTED CURRENT

In order to further clarify the effect of the angular frequency of the traveling wave on the motion of the double-headed Brownian motor, we calculate $\langle \dot{x} \rangle$ and $\langle \dot{y} \rangle$ as functions of the angular frequency $\omega$ for different values of the intensity of Gaussian white noise in Figs. 5(a) and (b), respectively. It is found that when the angular frequency is very small, the current $\langle \dot{x} \rangle$ is reversed as shown in the inset of Fig. 5(a). In Fig. 5(b), it is shown that the current $\langle \dot{y} \rangle$ is always negative. As it was predicted, it is preferable to move downward along the slope of the potential for the system in Fig. 3. Moreover, one can see that the absolute value $|\langle \dot{y} \rangle|$ increases monotonically with the angular frequency $\omega$ in Fig. 5(b), namely the negative current in the tilted potential increases with the increase of the tilt of the potential. For smaller angular frequencies of the traveling wave, $|\langle \dot{y} \rangle| > u$, so the current $\langle \dot{x} \rangle = \langle \dot{y} \rangle + u < 0$. At a certain angular frequency, $\langle \dot{y} \rangle + u = 0$. As the continuous increase of the angular frequency, the value of $\langle \dot{y} \rangle + u$ is turned to positive. So the current $\langle \dot{x} \rangle$ is reversed. It is shown again that the competition between the negative velocity of the center of mass induced by the rotational motion and the wave velocity causes the reversal of $\langle \dot{x} \rangle$.

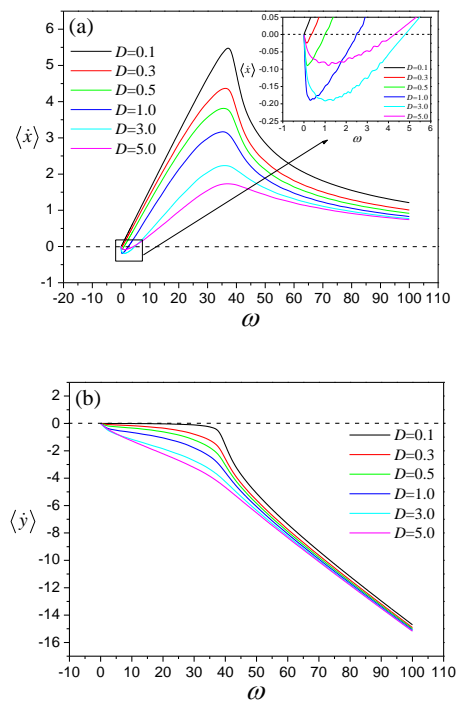

Fig. 5 The currents versus the angular frequency $\omega$ for different values of the noise intensity $D$. (a) $\langle \dot{x} \rangle$ versus $\omega$, and (b) $\langle \dot{y} \rangle$ versus $\omega$. Parameter values: $l=1.0$, $F=0.0$.

Furthermore, Figure 5(a) also shows that for larger angular frequencies, $\langle \dot{x} \rangle$ varies nonmonotonically with the angular frequency, and can be optimized at an appropriate angular frequency, which is in agreement with the result given in Ref. [1]. This can be interpreted in two ways. On the one hand, mathematically the change of the value of $\langle \dot{y} \rangle + u$ causes the non-monotonic variation of $\langle \dot{x} \rangle$. On the other hand, physically in terms of the traveling-wave potential, when the angular frequency is not too large, namely the wave does not travel very fast, the system is tightly coupled to the potential so that it can keep up with the wave. As increasing the angular frequency, the current $\langle \dot{x} \rangle$ also increases. However, when the angular frequency is large enough, the wave travels so fast that the coupling of the system to the potential becomes looser. In this case, the system can not keep up with the wave so that the current $\langle \dot{x} \rangle$ decreases with the increase of the angular frequency.

In Fig. 5(b), we can also find that for smaller noise intensities, the curve of $\langle \dot{y} \rangle$ versus $\omega$ is nonlinear. In the low-to-moderate range of the angular frequency, $\langle \dot{y} \rangle$ changes slowly with $\omega$, while fast for larger angular frequencies. This is mainly because that for the low-to-moderate angular frequencies and smaller noise intensities, the tilt of the potential is so small that the motor is trapped in the potential well for a longer time. Thus $\langle \dot{y} \rangle$ changes slowly with $\omega$. When the angular frequency is large enough, the current $\langle \dot{y} \rangle$ increases rapidly with $\omega$.

Besides, Figures 5(a) and (b) show for zero angular frequency, the initial values of $\langle \dot{x} \rangle$ and $\langle \dot{y} \rangle$ are all zero. This indicates that the traveling wave is the essential condition of the directed current because it induces the fundamental spatial symmetry breaking in the translational direction.

## C. EFFECT OF THE EXTERNAL DRIVING FORCE ON THE DIRECTED CURRENT

We consider a constant external driving force $F$ which is subject to the system in the rotational direction and can induce a continuous rotation of the double-headed Brownian motor. The currents $\langle \dot{x} \rangle$ and $\langle \dot{y} \rangle$ as functions of $F$ for different noise intensities are plotted in Fig. 6. First, we can clearly see that $\langle \dot{x} \rangle$ and $\langle \dot{y} \rangle$ are highly symmetrical for $F > 0$ and $F < 0$ due to the same effects of the positive and negative forces on the motion of the system in this model. So next, we only discuss the case $F > 0$.

Similar to Fig. 5(a), an obvious reversal of $\langle \dot{x} \rangle$ with increasing the force $F$ can be observed in Fig. 6(a). In addition, the current $\langle \dot{y} \rangle$ in the tilted potential is still negative, and exhibits a resonance-like phenomenon. Too small and too large forces are all unfavourable for the directed current $\langle \dot{y} \rangle$, and for a moderate force the directed motion of the system can be enhanced. The inset of Fig. 6(a) shows that for the same noise intensity, the difference between $\langle \dot{x} \rangle$ and $\langle \dot{y} \rangle$ is just the wave velocity $u$. The curve of $\langle \dot{x} \rangle$ against $F$ is equivalent to move the curve of $\langle \dot{y} \rangle$ against $F$ up the distance $u$ in the same coordinate system, which is consistent with the result from Eq. (13).

In Fig. 6(b), it is also found that for $D = 0$, there are two threshold values of the force $F_1$ and $F_2$. When $F < F_1 = 1$, too small force is not enough to make the system overcome the barrier of the tilted potential, indicating a null-directed motion. In this model, the minimum force to induce the directed motion should be equal to the barrier height of the potential in the absence of noise. Except the case $F < F_1$, the directed current in the tilted potential can not also be achieved for $F > F_2$. For the nonzero noise case, we can find that the thermal fluctuation can smooth the current. These results are similar to that in Ref. [39].

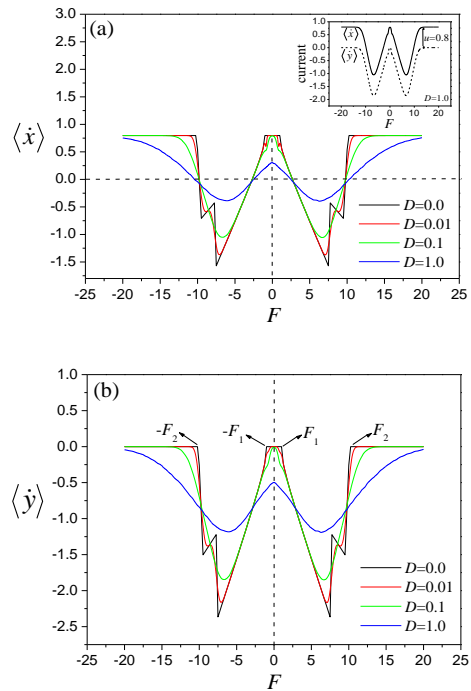

Fig. 6 The currents versus the external driving force $F$ for different values of the noise intensity $D$. (a) $\langle \dot{x} \rangle$ versus F, and (b) $\langle \dot{y} \rangle$ versus $F$. The inset of Fig. 6(a) shows the functions of $\langle \dot{x} \rangle$ and $\langle \dot{y} \rangle$ as $F$ for $D$=1.0. Parameter values: $l$=1.0, $\omega$=5.0.

## D. EFFECT OF THE ROD LENGTH ON THE DIRECTED CURRENT

To explore the effect of the spatial size of the system on directed motion, we calculate $\langle \dot{x} \rangle$ and $\langle \dot{y} \rangle$ as functions of the rod length $l$ in Figs. 7(a) and (b), respectively. One can find that the currents oscillate periodically with the rod length $l$. When $l/\lambda$ is approximately an integer, the currents can reach some local maxima. This is the result of the competition between two spatial scales, namely the rod length and the wavelength.

In Fig. 7(a), for a given angular frequency of the traveling wave, when $l$ is zero, the current $\langle \dot{x} \rangle$ has the maximum. This is because that from the point of view of the traveling-wave potential, the double-headed Brownian motor is regarded as a single particle and can travel with the wave. While in Fig. 7(b), the initial current is zero for $l$=0. The reason is that the directed current $\langle \dot{y} \rangle$ can not be achieved for a particle without rotational effect in a tilted potential in this model. Besides, in Fig. 7(a), the current reversal still appears for smaller angular frequencies, which is still caused by the change of the value of $\langle \dot{y} \rangle + u$.

## E. EFFECT OF THE NOISE ON THE DIRECTED CURRENT

We here consider that there is only the noise in the rotational direction, i.e., $F=0$. For different angular frequencies of the traveling wave, $\langle \dot{x} \rangle$ and $\langle \dot{y} \rangle$ versus the noise intensity $D$ are plotted in Figs. 8(a) and (b), respectively. In Fig. 8(b), the negative current $\langle \dot{y} \rangle$ in the tilted potential exhibits a typical stochastic resonance effect induced by Gaussian white noise, which is similar to that in other Brownian motors in a tilted potential [53].

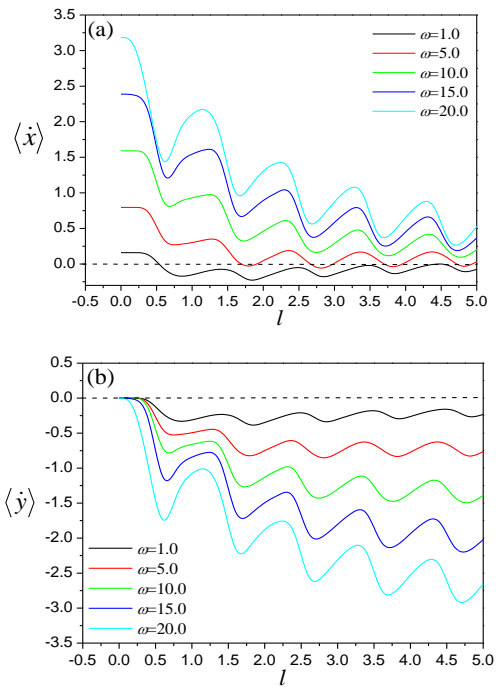

Fig. 7 The currents versus the rod length $l$ for different values of the angular frequency $\omega$. (a) $\langle \dot{x} \rangle$ versus $l$, and (b) $\langle \dot{y} \rangle$ versus $l$. Parameter values: $D=1.0$, $F=0.0$.

For a very weak noise, in the absence of the external force the rotational motion of the motor is very weak so that the center of mass moves very slowly. In this case, the residence time of the motor system in the potential well increases, and so the directed current $\langle \dot{y} \rangle$ is very small. For the very strong noise, the orderly directed rotation of the motor is destroyed due to the strong thermal fluctuation. For this case, the directed current induced by the rotational motion is also very weak. Therefore, a moderate noise intensity can enhance the directed motion of the system. In addition, we can also see that for a fixed noise intensity, the directed current $\langle \dot{y} \rangle$ increases with increasing the tilt of the potential (i.e., the wave velocity). This is because that it is much easier for the motor system to across a smaller potential barrier along the slope for a larger tilt. These results agree with that in Fig. 5(b).

Figure 8(a) shows that for smaller angular frequencies, a current reversal appears. The reason still comes from the competition between the negative current $\langle \dot{y} \rangle$ and the wave velocity $u$. The current $\langle \dot{x} \rangle$ with the noise intensity $D$ shows the reverse behavior of the typical stochastic resonance phenomenon. It is easy to understand according to the previous analysis. We have well known that the current $\langle \dot{x} \rangle$ is the sum of the negative current $\langle \dot{y} \rangle$ and the wave velocity $u$. As shown in Fig. 8(b), for very small and large noise intensities, the negative current $\langle \dot{y} \rangle$ is restrained, and enhanced for moderate noise intensities. On the contrary, the current $\langle \dot{x} \rangle$ is promoted for very small and large noise intensities, and restrained for moderate noise intensities according to Eq. (13). Hence, when $\langle \dot{y} \rangle$ presents a typical stochastic resonance, $\langle \dot{x} \rangle$ shows the reverse of stochastic resonance.

Besides, in Figs. 8(a) and (b), we also see that for the same angular frequency and noise intensity, the difference between $\langle \dot{x} \rangle$ and $\langle \dot{y} \rangle$ is still $u$ as shown in the inset of Fig. 8(a).

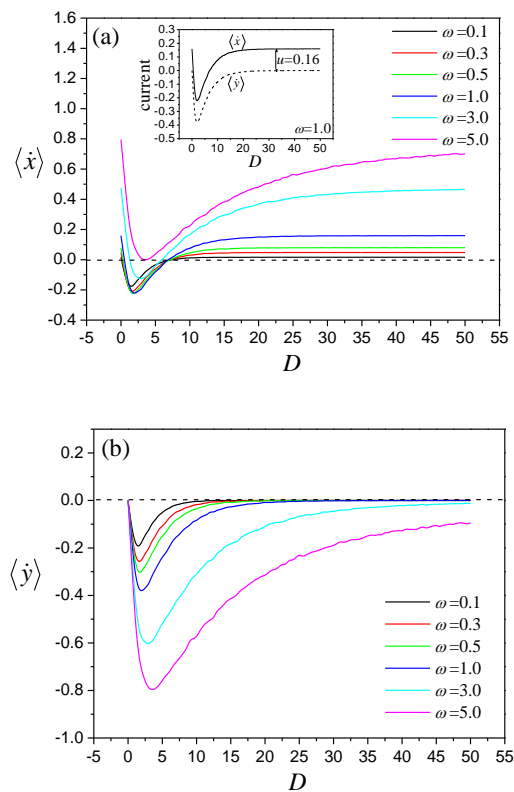

Fig. 8 The currents versus the noise intensity $D$ for different values of the angular frequency $\omega$. (a) $\langle \dot{x} \rangle$ versus $D$, and (b) $\langle \dot{y} \rangle$ versus $D$. The inset of Fig. 8(a) shows the functions of $\langle \dot{x} \rangle$ and $\langle \dot{y} \rangle$ as $D$ for $\omega=1.0$. Parameter values: $l=1.0$, $F=0.0$.

### F. EFFECT OF A NON-GAUSSIAN NOISE ON THE DIRECTED CURRENT

In the above investigations, we focus on the Gaussian white noise in the rotational direction. To explore the impact of a non-Gaussian noise on the directed transport, we replace the Gaussian white noise with the white α-stable symmetric Lévy noise $\xi_\alpha(t)$. Different from the Gaussian noise, for the Lévy noise, its distribution has a fat tail characterized by $|\xi|^{-1-\alpha}$, and the characteristic function of the distribution is given by $p(k) = \exp(-\sigma|k|^\alpha)$. The parameter α ($\alpha \in (0, 2]$) denotes the stability index of the distribution. The smaller α is, the more obvious the fat-tailed characteristic is. The case $\alpha=1$ corresponds to the Cauchy distribution and the limit case $\alpha=2$ corresponds to the Gaussian distribution. In this work, we are mainly concerned with the directed current of the motor, so $\alpha > 1$ is restricted so that the first moment of the Lévy distribution is finite. The parameter $\sigma$ ($\sigma > 0$) is the noise intensity of Lévy noise. The random number $\xi_\alpha(t)$ that meets the Lévy distribution is generated by Chambers–Mallows–Stuck (CMS) algorithm [60, 61]:

$$\xi_\alpha = \left(\frac{-\log\mu\cos\varphi}{\cos[(1-\mu)\varphi]}\right)^{1-1/\alpha} \frac{\sin(\alpha\varphi)}{\cos\varphi}, \quad (20)$$

where $\varphi = \pi(v - \frac{1}{2})$. $\mu$ and $v$ are independent uniform random numbers distributed on (0,1).

Adopting the same algorithm and calculating parameters as that in the above sections, we calculate the directed currents as functions of the angular frequency $\omega$ of the traveling wave, external force $F$ and the noise intensity $\sigma$ for different values of the stability index of Lévy noise. Note that the case $\alpha=2.0$ corresponding to the Gaussian white noise is considered to make a comparison with Lévy noises.

Figures 9(a) and (b) show the currents $\langle\dot{x}\rangle$ and $\langle\dot{y}\rangle$ as functions of the angular frequency of the traveling wave. We can see that the directed currents show similar dependence on the angular frequency for both cases of Lévy and Gaussian noises. In Fig. 9(b), the negative current $\langle\dot{y}\rangle$ is sensitive to the stability index $\alpha$ of the Lévy noise for a low-to-moderate range of the angular frequency, namely for not too large tilt of the potential. The smaller the stability index $\alpha$ of the noise is, the smaller the negative current is, which is shown in the inset of Fig. 9(b). This is because that for not too large tilts of the potential, if noise exists only in the rotational direction, it is hard to cause an orderly directed rotation of the system for Lévy noise with a smaller stability index. Thus, in this case the Gaussian noise ($\alpha=2$) can promote the negative current more effectively induced by the rotational motion than the Lévy noise. A similar result can be found in Ref. [55]. For very large angular frequency (i.e., very large tilt of the potential), the effect of the potential is much stronger than that of the noise on the directed current so that the sensitivity of the directed current to the stability index decreases. In Fig. 9(a), the effect of the stability index on the current $\langle\dot{x}\rangle$ is the reverse of that on $\langle\dot{y}\rangle$, which can be also reflected from Eq. (13).

We also calculate the directed currents $\langle\dot{x}\rangle$ and $\langle\dot{y}\rangle$ as functions of the external force $F$ for different values of the stability index $\alpha$ of the Lévy noise in Figs. 10(a) and (b). The curves of the directed currents versus the external force for both cases of Lévy and Gaussian noises have the same shapes. When the force is zero, the effects of the stability index on the directed currents agree with that in Fig. 9. In Fig. 10(b), the current $\langle\dot{y}\rangle$ versus the force still shows a resonance-like effect. For an appropriate force, $\langle\dot{y}\rangle$ can reach its maximum. Moreover, the smaller the stability index is, the greater the peak current is. Here, it is indicated that the directed current $\langle\dot{y}\rangle$ can be optimized by modulating the appropriate parameters. In Fig. 10(a), the current $\langle\dot{x}\rangle$ against $F$ is also equivalent to move the curve of $\langle\dot{y}\rangle$ against $F$ up the distance $u$ in the same coordinate system.

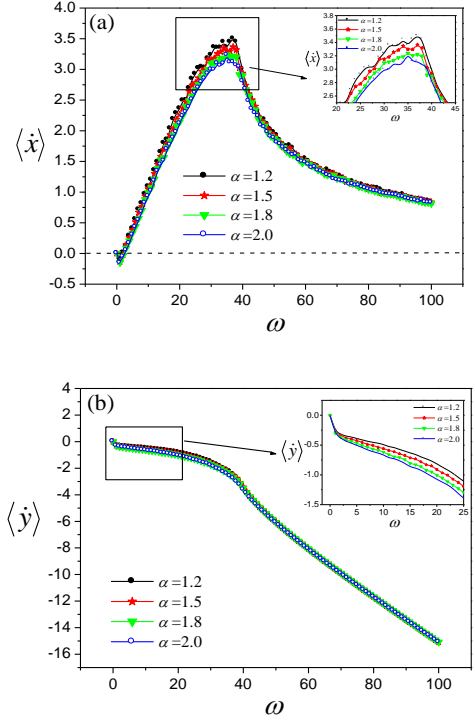

Fig. 9 The currents versus the angular frequency $\omega$ for different values of the stability index $\alpha$. (a) $\langle\dot{x}\rangle$ versus $\omega$, and (b) $\langle\dot{y}\rangle$ versus $\omega$. Parameter values: $l=1.0$, $F=0.0$, $\sigma=1.0$.

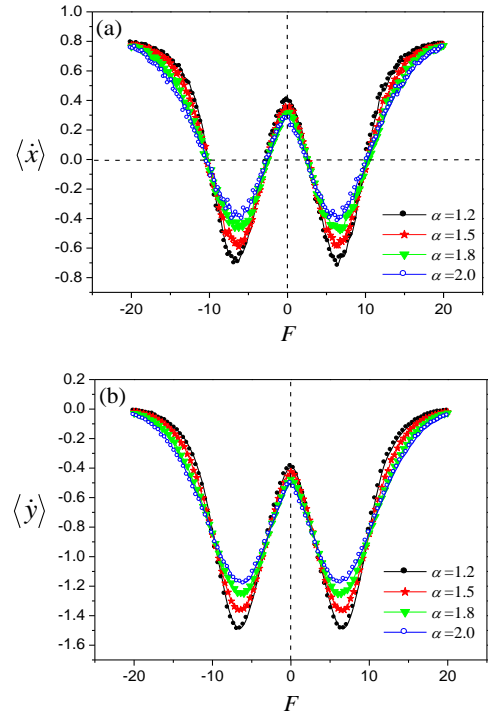

Fig. 10 The currents versus the external driving force F for different values of the stability index $\alpha$. (a) $\langle\dot{x}\rangle$ versus $F$, and (b) $\langle\dot{y}\rangle$ versus $F$. Parameter values: $l=1.0$, $\omega=5.0$, $\sigma=1.0$.

Figures 11(a) and (b) show the effects of the noise intensity $\sigma$ on the directed currents $\langle\dot{x}\rangle$ and $\langle\dot{y}\rangle$. Similar to the case of the Gaussian noise, the negative current $\langle\dot{y}\rangle$ also shows the typical stochastic resonance under the Lévy noise as shown in Fig. 11(b). For an appropriate noise intensity, the directed current can achieve a maximum which is seen in other models with Lévy noises [55, 57]. In addition, we also find that in different ranges of the noise intensity, the stability index of the noise has also different effects on the current. For smaller noise intensities, the current $\langle \dot{y} \rangle$ increases with increasing the stability index, and decreases with the increase of the stability index for larger noise intensities, which is also similar to the results in Ref. [55]. In Fig. 11(a), in the traveling-wave potential, an opposite effect of the stability index on the current $\langle \dot{x} \rangle$ is shown.

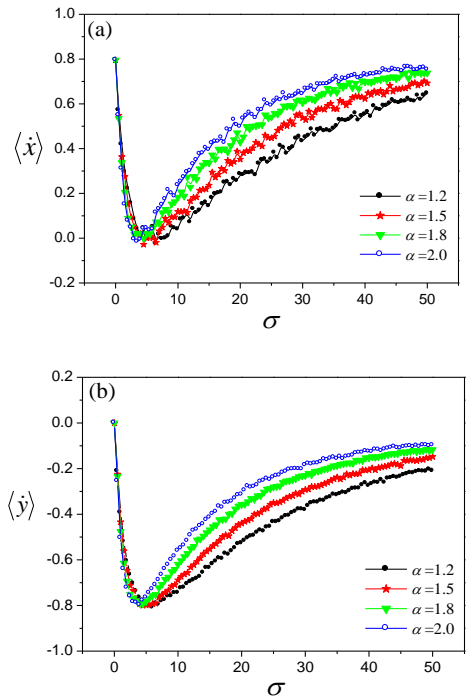

Fig. 11 The currents versus the noise intensity $\sigma$ for different values of the stability index $\alpha$. (a) $\langle \dot{x} \rangle$ versus $\sigma$, and (b) $\langle \dot{y} \rangle$ versus $\sigma$. Parameter values: $l=1.0$, $\omega=5.0$, $F=0.0$.

## IV. Conclusions

In this paper, we investigate the rotation-translation coupling dynamics of a double-headed Brownian motor in a traveling-wave potential. The noise and an external constant driving force are subject to the system only in the rotational direction. Abundant directed transport characteristics of the motor system are revealed.

We first find that the essential condition for the directed motion of the motor system is the presence of the traveling wave because it provides the fundamental spatial asymmetry in the translational direction. Then, the current reversal in the traveling-wave potential generally appears. By transforming the dynamical equation in traveling-wave potential into that in a tilted potential, we analyze the mechanism of the current reversal. It is the competition between the negative velocity of the center of mass induced by the rotational motion and the positive wave velocity that causes the reversal of the current.

Moreover, in the presence of the Gaussian white noise, the current in the tilted potential exhibits a typical stochastic resonance effect, and an appropriate noise intensity can enhance the current. The external driving force has also a resonance-like effect on the current. The spatial size of the motor, namely the rod length, also affects the directed transport. Under some characteristic values of the rod length, the current can achieve its extremum. However, the current in the traveling-wave potential has the reverse behaviors of that in the tilted potential.

Finally, when one uses a non-Gaussian Lévy noise to replace the Gaussian white noise, the effects of various parameters on the directed current are explored. The results show that the currents show similar dependence on the parameters for both cases of Lévy and Gaussian noises. In addition, the currents obviously depend on the stability index of the Lévy noise under certain conditions. For zero force and small noise intensities, the directed current in the tilted potential increases with increasing the stability index. On the other hand, for nonzero force and large noise intensities, the directed current in the tilted potential possibly decreases with increasing the stability index.

In conclusion, in this work, we develop a rotation-translation model of a double-headed Brownian motor, and provide a valuable way of analyzing the directed transport in traveling-wave potential. In future studies, we will further explore some relevant issues such as the rotation-translation coupling dynamics in a randomly switching potential and some other effects induced by the Lévy noise (e.g., noise enhanced stability, stochastic resonant activation).

## Acknowledgements

This work has been supported by the Key project of Beijing Institute of Graphic Communication, Beijing, China (No. Ea201702), the International Ability Improvement Project of Teaching Staff of Beijing Institute of Graphic Communication, Beijing, China (No.12000400001), the National Natural Science Foundation of China (Grant No. 11875135), Quanzhou Scientific and Technological Foundation (No. 2018C085R), and the Innovation Teams in Functional materials and structural mechanics of Hebei University of Architecture (No. TD202011).